# Modeling and Simulation of Spin Transfer Torque Generated at Topological Insulator/Ferromagnetic Heterostructure

A. K. Reza,[1,a)] X. Fong,[1,b)] Z. A. Azim [1,c)] and K. Roy[1,d)]

[1]*Electrical and Computer Engineering, Purdue University, West Lafayette, Indiana, 47905, USA*

Topological Insulator (TI) has recently emerged as an attractive candidate for possible application to spintronic circuits because of its strong spin orbit coupling. TIs are unique materials that have an insulating bulk but conducting surface states due to band inversion and these surface states are protected by time reversal symmetry. In this paper, we propose a physics-based spin dynamics simulation framework for TI/Ferromagnet (TI/FM) bilayer heterostructures that is able to capture the electronic band structure of a TI while calculating the electron and spin transport properties. Our model differs from TI/FM models proposed in the literature in that it is able to account for the 3D band structure of TIs and the effect of exchange coupling and external magnetic field on the band structure. Our proposed approach uses 2D surface Hamiltonian for TIs that includes all necessary features for spin transport calculations so as to properly model the characteristics of a TI/FM heterostructure. Using this Hamiltonian and appropriate parameters, we show that the effect of quantum confinement and exchange coupling are successfully captured in the calculated surface band structure compared with the quantum well band diagram of a 3D TI, and matches well with experimental data reported in the literature. We then show how this calibrated Hamiltonian is used with the self-consistent non equilibrium Green's functions (NEGF) formalism to determine the charge and spin transport in TI/FM bilayer heterostructures. Our calculations agree well with experimental data and capture the unique features of a TI/FM heterostructure such as high spin Hall angle, high spin conductivity etc. Finally, we show how the results obtained from NEGF calculations may be incorporated into the Landau–Lifshitz–Gilbert–Slonczewski (LLGS) formulation to simulate the magnetization dynamics of an FM layer sitting on top of a TI.

___________________________

[a)] areza@purdue.edu      [b)] xfong@ecn.purdue.edu      [c)] zazim@purdue.edu      [d)] kaushik@purdue.edu

**I. INTRODUCTION:**

Topological Insulators (TI) are a new class of materials which are characterized by unique quantum-mechanical properties due to their unusual surface states. Although the bulk of a TI is insulating, the surface is conducting due to band inversion at the surface. This band inversion is a consequence of the high spin-orbit coupling (SOC) in TIs which causes the valence band and the conduction band to touch each other at the interface [1]. This strong SOC also enables a TI to manipulate the magnetization of an adjacent ferromagnet (FM) layer by generating a high Spin-Orbit Torque (SOT) through the Rashba-Edelstein effect [2] [3] [4] [5]. This torque is further amplified by helical locking of the relative orientation of spin and momentum at the conducting surface states [2] [6] [7]. Recent experiments [2] [5] have clearly demonstrated this strong SOT acting on the FM layer in a TI/FM heterostructure. Moreover, the unique properties of a TI give rise to an unusually high spin Hall angle [2]. While spin Hall materials such as Ta, W and other heavy metals show spin Hall angles less than 0.3, TI has been experimentally found to exhibit much higher spin Hall angle (~1.1) [8]. This efficiency in generating spin currents with lower charge current injection has attracted extensive research interest in TI/FM heterostructures. Consequently, a lot of effort has been made to model the behavior of a TI/FM heterostructure [2] [4] [6]. But none of them consider the overall simulation framework for calculating the spin transport and the magnetization dynamics of the FM layer.

In this work, for analyzing the performance of a TI based memory device, we are going to propose a complete simulation framework that is computationally inexpensive yet shows good match with the experimental result. Step by step modelling detail is shown in fig. 1. We will first design a Hamiltonian that includes the significant aspects of TI/FM heterostructure that affect the charge and spin transport. We have formulated a 2D surface Hamiltonian for a TI structure by considering the quantum confinement effect, the position of the Fermi level, Dirac cone, Rashba effect, exchange coupling energy with the adjacent FM layer and the effect of external magnetic field on the band structure. We compared the band structure resulting from our proposed Hamiltonian with the band diagram from a standard 4x4 k.p [9] Hamiltonian of TI. The band structure we obtain shows excellent match with 3D TI band-structures. Next, we calculate the electrical transport characteristics by applying our Hamiltonian to standard non equilibrium Green's functions (NEGF) formalism of quantum transport. We determine the charge current, the spin current, the charge conductivity and the in-plane and out-of-plane spin conductivity through self-consistent NEGF simulations. From the ratio of the obtained charge current and spin current, we calculate the spin Hall angle and find good match with experimental data. Finally, we analyzed the magnetization dynamics of a TI/FM heterostructure by incorporating the NEGF calculations into the Landau–Lifshitz–Gilbert–Slonczewski (LLGS) magnetization dynamics model.

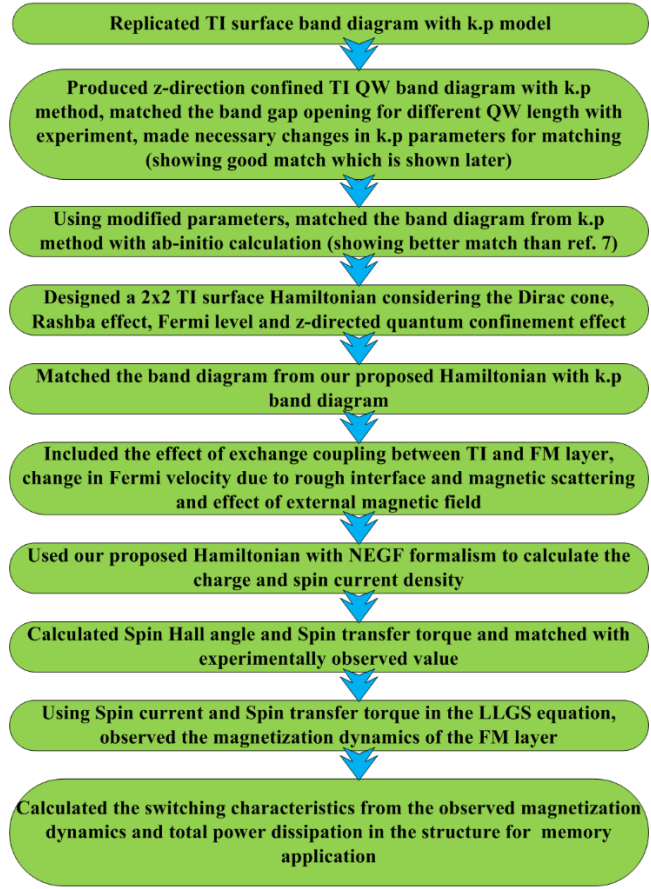

Fig 1: Flow diagram of the modelling detail

The rest of the article is organized as follows. In section II, we present the details of our proposed model, including the NEGF transport method and LLGS magnetization dynamics. In sections III(A) and III(B), we apply our model and analyze two different TI/FM heterostructures (Cr doped $(Bi_{0.5}Sb_{0.5})_2Te_3/(Bi_{0.5}Sb_{0.5})_2Te_3$ heterostructure and Permalloy $(Ni_{0.81}Fe_{0.19})$ / $Bi_2Se_3$ heterostructure) and check the consistency of our results with experimental observations [2] [5] [8]. In section III(C) we show how a magnetic memory may be designed using TI/Ferromagnet heterostructures. Section IV concludes the article.

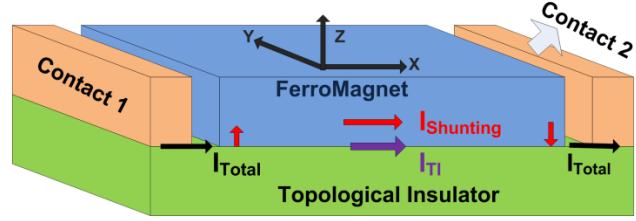

Fig. 2: Schematic drawing of a TI/FM heterostructure where the total charge current is flowing through the TI surface along the x axis from Contact 1 to Contact 2. One part of the charge current will shunt through the top FM layer and the rest will flow through the TI surface. Charge current is spin polarized in perpendicular directions i.e., along y and z axis. Accumulated spin in the TI surface diffuses in the FM layer creating the torque.

## II. PROPOSED MODEL OF TOPOLOGICAL INSULATOR / FERROMAGNET HETEROSTRUCTURE

### A. Band Structure of TI

The first step in developing our numerical model is to replicate the band structure of a TI using a simple k.p model [9] for couple of reasons. First the k.p method exhibits excellent agreement with experimental results (shown later) for predicting the quantum confinement gap opening in TI quantum well band structure. Therefor we can use the k.p method to observe the quantum confinement effect on band structure of TI quantum well of different length. Secondly we can benchmark the band diagram from our proposed Hamiltonian with it. The band structure of a TI is unique because the bulk is insulating but the surface is conducting due to band inversion at the surface. TI materials have very high spin orbit coupling (SOC) and this causes the conduction and valence bands to touch each other in the surface at the $\Gamma$ point (center of the Brillouin zone in reciprocal lattice, where the wave vector k=0), thereby creating band inversion (in fig. 3). Electrical current primarily flows at the surface of a TI due to this band inversion at the $\Gamma$ point. Therefore, the band diagram and transport properties of a TI can be characterized by the physics near the $\Gamma$ point. The standard k.p model is reported to be accurate for modeling the band structure near the $\Gamma$ point; [10] hence; we use it to benchmark our TI model. We first develop a simple 2×2 Hamiltonian to model the TI surface by considering important physical properties of the surface. We consider the surface Fermi velocity, Fermi level of the 3D band diagram, the quantum confinement effect and the Rashba-Edelstein effect. Next we apply this Hamiltonian to extract the band structure and compare it with the band diagram from a standard k.p Hamiltonian of TI [1] [7]. The k.p model we consider here is a 4x4 k.p model for 3D TI [7] which includes important symmetry properties such as time reversal symmetry, inversion symmetry and three-fold rotation symmetry along z-axis. The k.p dispersion relation in 3D TI is computed by considering the four low lying states: $|P1_z^+ \uparrow\rangle$, $|P1_z^+ \downarrow\rangle$, $|P2_z^- \uparrow\rangle$ and $|P2_z^- \downarrow\rangle$ which are closest to the Fermi level [1] (schematically represented in FIG. 2). The dispersion relation for a finite wave vector **k** is expressed by the following Hamiltonian [1]:

$$H(\mathbf{k}) = \epsilon_0(\mathbf{k})I_{4\times4} + \begin{pmatrix} M(\mathbf{k}) & A_1 k_z & 0 & A_2 k_- \\ A_1 k_z & -M(\mathbf{k}) & A_2 k_- & 0 \\ 0 & A_2 k_+ & M(\mathbf{k}) & -A_1 k_z \\ A_2 k_+ & 0 & -A_1 k_z & -M(\mathbf{k}) \end{pmatrix} \quad (1)$$

Here, $k_\pm = k_x \pm ik_y$, $\epsilon_0(\mathbf{k}) = C + D_1 k_z^2 + D_2(k_x^2 + k_y^2)$ and $M(\mathbf{k}) = M - B_1 k_z^2 - B_2(k_x^2 + k_y^2)$ [1]. With the parameters given in ref. [7], the band structure of Bi$_2$Se$_3$ obtained from this Hamiltonian matches well with the *ab initio* calculations. Moreover, the Quantum Well (QW) bandgap opening shows good agreement with experimental observations [11] [12].

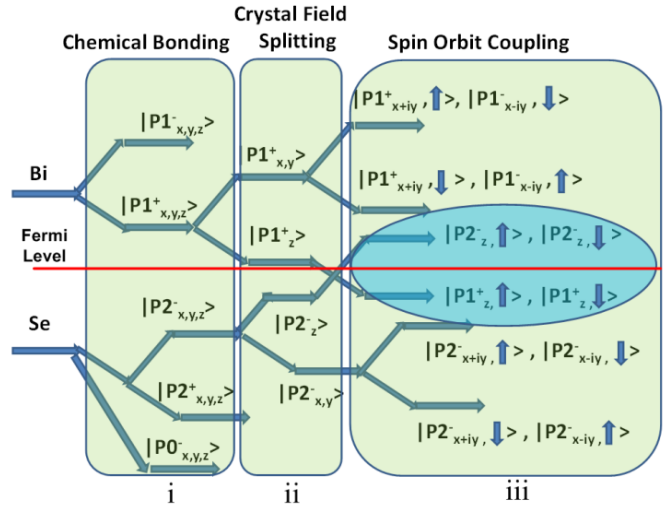

Fig. 3: Schematic representation energy level splitting of 'p' orbitals in Bi$_2$Se$_3$ at the Γ point due to (i) chemical bonding between Bi and Se, (ii) Crystal field splitting and (iii) spin orbit coupling [1].

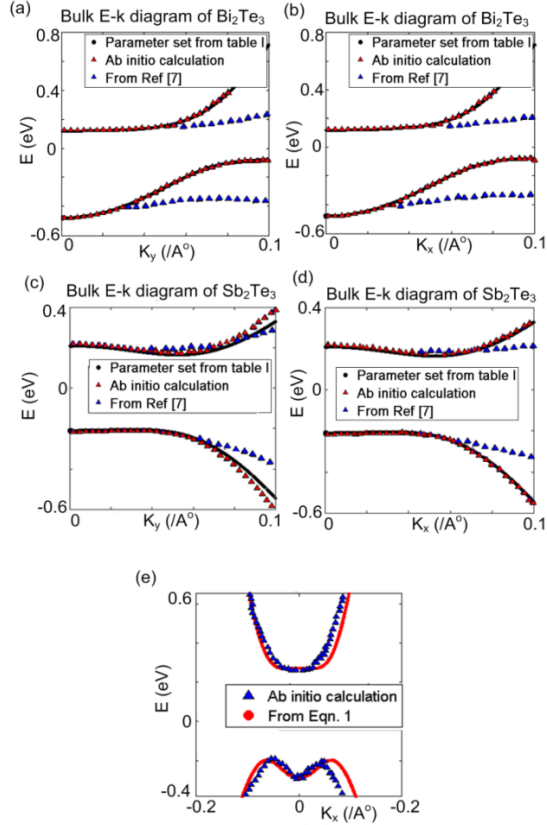

Fig. 4: Bulk band diagram comparison of (a) $Bi_2Te_3$ along $k_y$ (b) $Bi_2Te_3$ along $k_x$ (c) $Sb_2Te_3$ along $k_y$ (d) $Sb_2Te_3$ along $k_x$ and (e) $Bi_2Se_3$ along $k_x$. (*Ab initio* calculations are shown in reference [7]).

Although, this model is adequate for $Bi_2Se_3$ with parameters reported in ref. [7], we observe the inadequacies of the model when we apply the parameters to $Bi_2Te_3$ and $Sb_2Te_3$. In fig. 4, we show the comparison of the bulk E-k dispersion diagram using the parameters in ref. [7] with *ab initio* calculations around the Γ point. We notice that the k.p band structure of $Bi_2Te_3$ and $Sb_2Te_3$ do not sufficiently match with the *ab initio* calculations.

TABLE I. Parameters for $Bi_2Se_3$, $Bi_2Te_3$ and $Sb_2Te_3$

| Parameters | $Bi_2Se_3$ parameters from ref [7] | $Bi_2Te_3$ parameters from ref [7] | Modified parameters for $Bi_2Te_3$ | $Sb_2Te_3$ parameters from ref [7] | Modified parameters for $Sb_2Te_3$ |
|---|---|---|---|---|---|
| M | -0.28 eV | -0.3 eV | -0.3 eV | -0.22 eV | -0.22 eV |
| A1 | 2.26 eV- Å | 0.3 eV- Å | 0.3 eV- Å | 0.84 eV- Å | 0.84 eV- Å |
| A2 | 3.33 eV- Å | 2.87 eV- Å | 2.303 eV- Å | 3.4 eV- Å | 3.4 eV- Å |
| B1 | -6.86 eV - Å² | -2.79 eV - Å² | - 3.79 eV - Å² | -19.64 eV - Å² | - 19.64 eV - Å² |
| B2 | -44.5 eV - Å² | -57.38 eV - Å² | -57.38 eV - Å² | -48.51 eV - Å² | -48.51 eV - Å² |
| C | -0.0083 eV | -0.18 eV | -0.18 eV | 0.001 eV | 0.001 eV |
| D1 | 5.74 eV - Å² | 6.55 eV - Å² | 0.3 eV - Å² | -12.39 eV - Å² | -15.39 eV - Å² |
| D2 | 30.4 eV - Å² | 49.68 eV - Å² | 49.68 eV - Å² | -10.78 eV - Å² | -10.78 eV - Å² |

In addition, when we construct a z-confined thin quantum well band structure, we observe that the conduction and valence bands overlap, which is incorrect and this is a direct consequence of the inaccurate parameters. We also vary the quantum confinement length i.e. the quantum well thickness from 1 nm to 6 nm and measure the bandgap opening up at the $\Gamma$ point and find that it does not show any match with the experimental data [11] [12]. Therefore, we modified the parameter set of ref. [7] in a trial and error method until we got better match with the *ab initio* calculations for $Bi_2Te_3$ and $Sb_2Te_3$. The modified parameters are listed in table I. Using these modified parameters, we find excellent match with *ab initio* calculations as shown in fig. 4. Moreover, with these modified parameters, we get a good match with experimental data when we calculate the band gap opening due to the quantum confinement in the z-direction (details in section III).

## B. Surface Hamiltonian Modeling Including External Perturbations

As discussed earlier, electrical current primarily flows through the surface states of a TI which are also protected by time reversal symmetry [1] [13]. Hence, for electrical transport calculations, it is sufficient to model the Hamiltonian at the surface in the vicinity of the $\Gamma$ point. We developed a surface Hamiltonian model including the quantum confinement effect in the z-direction and considered external perturbations such as external magnetic field and exchange coupling with an adjacent Ferromagnet. We later use this surface Hamiltonian in conjunction with the standard NEGF method to calculate the electrical transport characteristics in a TI channel.

Let us first discuss the Hamiltonian for modeling the TI surface without any perturbations. The top and bottom surfaces of a TI can be modeled by a simple Hamiltonian as follows [14]:

$$H_D(\mathbf{k}) = \begin{pmatrix} 0 & iv_F k_- & m_k^* & 0 \\ -iv_F k_+ & 0 & 0 & m_k^* \\ m_k & 0 & 0 & -iv_F k_- \\ 0 & m_k & iv_F k_+ & 0 \end{pmatrix} \quad (2)$$

Here, $v_F$ is the Fermi velocity at the TI surface. The basis of this Hamiltonian are $|t \uparrow\rangle, |t \downarrow\rangle, |b \uparrow\rangle$ and $|b \downarrow\rangle$ where t and b denote the top and the bottom surfaces respectively, and $m_k$ represents the tunneling effect between the top and the bottom surfaces. It can be seen from fig. 2 that the spin-transfer torque acting on the ferromagnetic layer arises due to the current flow through the top surface. Therefore, only the Hamiltonian for the top surface is required for transport calculations and the tunneling between top and bottom surface can be ignored. Hence, the effective Hamiltonian can be written as:

$$H_D^{eff}(k) = \begin{pmatrix} 0 & iv_F k_- \\ -iv_F k_+ & 0 \end{pmatrix} \tag{3}$$

Note that, this is a Dirac Hamiltonian of the form, $H = v_F(\hat{z} \times \vec{\sigma}) \cdot \vec{k}$ ($\vec{\sigma}$ is the Pauli spin matrix, $k = k_x + k_y$) and only the Dirac type surface states are modeled. Consequently, this Hamiltonian captures the Dirac cone at the Γ point of the TI. The Hamiltonian is then modified in the following way to include the Fermi level:

$$H_D^{eff}(k) = \begin{pmatrix} \mu & iv_F k_- \\ -iv_F k_+ & \mu \end{pmatrix} \tag{4}$$

Here $\mu$ denotes the position of the Fermi level. Quantum confinement in the z-direction opens up a bandgap at the Dirac point [2] which is represented as $\Delta_{conf}$. When a ferromagnetic layer (FM) is placed on top of a TI surface, the exchange coupling energy needs to be considered as well. This exchange coupling arises as the spin of the TI surface state is coupled with the FM moment [15]. This results in increasing the bandgap opening at the Dirac point and is represented by $\Delta_{ex}$. This bandgap opening is present when the top FM layer has an out-of-plane i.e., z-directed magnetization. But for in plane magnetization, though strong exchange coupling may exist at the surface, no bandgap opening has been observed [16][17]. The combined bandgap opening is represented as: $\Delta_{gap} = \Delta_{conf} + \Delta_{ex}$. It can be included in the Hamiltonian as follows:

$$H_D^{eff}(k) = \begin{pmatrix} \mu + \Delta_{gap} & iv_F k_- \\ -iv_F k_+ & \mu - \Delta_{gap} \end{pmatrix} \tag{5}$$

Here one thing is worth mentioning that except bandgap opening the exchange coupling will have some other effects on the interface due to orbital overlapping between TI and FM layer. Therefore a 20 band $sp^3d^5s^*$ tight binding model will give more precise answer. But as we are mainly interested in transport calculation with less computational expense, we have tried to model the major effects that affect the surface transport. In the rough surface there is a strong suppression of transport channel [18]. Again there will be strong magnetic scattering in the surface which we have not considered in this model. This magnetic scattering causes weak anti-localization in the topological surface state [18]. As we have not considered this scattering our simulation shows a little higher conductivity than the experimental result especially where the surface is very rough like Permalloy/$Bi_2Se_3$ interface. Also due to orbital overlapping the surface Fermi velocity gets modified and this is included via off-diagonal parameter correction which is discussed later.

When electrical current flows through the surface of a TI, two types of spin-orbit torques is applied on the FM layer – the field like spin-orbit torque and the spin-transfer like spin-orbit torque [5]. The field like torque is proportional to the exchange coupling [15]. Since the surface Hamiltonian of eqn. (5) includes the effect of exchange coupling energy, it can be used to calculate the field like torque. It has been experimentally observed that the field-like torque is usually several orders of magnitude lower than the Slonczewski spin transfer torque [5]. This Slonczewski spin transfer torque arises due to the Rashba Spin-Orbit coupling (SOC) in two dimensional electron gas. Rashba Spin-Orbit coupling can be modeled as $H_{Rashba} = \frac{p^2}{2m} - \frac{\lambda}{\hbar} \vec{S} \cdot (\hat{z} \times \vec{p})$ [19] where p is the momentum, m is the effective mass, $\lambda$ is the Rashba coupling parameter and $\vec{S}$ is the Pauli matrices. As shown in fig 2 charge current is flowing along x-axis. Due to spin-momentum helical locking there is spin accumulation in y direction. Therefore spin angular momentum will flow in the FM layer and exerts spin orbit torque [2]. In our TI model, the perturbed surface state Hamiltonian is used to capture the Rashba Spin-Orbit coupling as shown below:

$$H^{eff}_{surface}(k) = \begin{pmatrix} B_1(k_x^2 + k_y^2) + \mu + \Delta_{gap} & i(v_F k_- - \lambda k_-) \\ -i(v_F k_- - \lambda k_-) & -B_2(k_x^2 + k_y^2) + \mu - \Delta_{gap} \end{pmatrix} \quad (6)$$

Where, $B_1$ and $B_2$ are two fitting parameters for modeling the Rashba splitting of 2D surface electron gas with the unit of eV-Ang$^2$. Exact values of $B_1$ and $B_2$ are determined by matching the surface band diagram with the z-confined 3D Quantum Well (QW) band structure. It has been observed that values of $B_1$ and $B_2$ do not depend on the confinement length in z-direction because the confinement effect has already been captured by the parameter, $\Delta_{gap}$.

Additionally, the interfacial states between the TI and FM affect the conductivity and the total spin current at the TI surface. For example, in the Permalloy/Bi$_2$Se$_3$ heterostructure, Permalloy is a face centered cubic structure (lattice constant ~0.355 nm) [20] whereas Bi$_2$Se$_3$ is a rhombohedral crystal with hexagonal supercell (lattice constants: a = 0.4318 nm and c = 2.864 nm) [21]. Due to the mismatch in lattice constants, we need to rewrite the Hamiltonian of the TI surface to include the change of Fermi velocity at the surface. Therefore due to inclusion of Rashba effect and modified Fermi velocity the Hamiltonian can be written in the following way:

$$H^{eff}_{surface}(k) = \begin{pmatrix} B_1(k_x^2 + k_y^2) + \mu + \Delta_{gap} & A k_- \\ A k_+ & -B_2(k_x^2 + k_y^2) + \mu - \Delta_{gap} \end{pmatrix} \quad (7)$$

Here A is another fitting constant with a unit of eV-Ang. A includes the off-diagonal Rashba effect and the modified Fermi velocity.

External magnetic fields can also act as a strong external perturbation on the TI surface states. Therefore, it is important to include the effect of it. The applied external magnetic field in any arbitrary direction can be represented by: $\vec{B_{ext}} = B(\sin\theta\cos\varphi\hat{x} + \sin\theta\sin\varphi\hat{y} + \cos\theta\hat{z})$. If $\vec{A}$ is the vector potential of the magnetic field defined such that, $\vec{B_{ext}} = \nabla \times \vec{A}$; then the effect of magnetic field can be included in the Hamiltonian by Peierls phase substitution, $k_{new} = k - \frac{e}{\hbar c}A$ [22]. Now, if the external constant magnetic field (B) is in-plane i.e. in the x-y plane (fig. 2) producing an angle of $\varphi$ with x-axis, the vector magnetic potential or the Landau Gauge [22] can be expressed as $\vec{A} = (Bz\sin\varphi, -Bz\cos\varphi, 0)$ where z is the confined length along z-axis. The wave vectors $k_x$ and $k_y$ are then transformed to: $k_{x,new} = k_x - \frac{e}{\hbar}Bz\sin\varphi$ and $k_{y,new} = k_y + \frac{e}{\hbar}Bz\cos\varphi$. For a particular confinement length (z) and constant magnetic field (B), the factor '$\frac{e}{\hbar}Bz$' is a constant. Hence, the Landau Gauge depends only on the angle $\varphi$. Again, if the external magnetic field is in the x-direction then, $\vec{A} = (0, -Bz, 0)$ and the wave vector $k_x$ remains unchanged and $k_{y,new} = k_y + \frac{e}{\hbar}Bz$. Similarly, for y-directed magnetic fields, the Landau gauge becomes: $\vec{A} = (Bz, 0, 0)$, $k_y$ remains unchanged while, $k_{x,new} = k_x - \frac{e}{\hbar}Bz$. Therefore, for an in-plane magnetic field the Hamiltonian can be written as follows:

$$H_{surface}^{eff}(k) = \begin{pmatrix} B_1(k_{x,new}^2 + k_{y,new}^2) + \mu + \Delta_{gap} & Ak_- \\ Ak_+ & -B_2(k_{x,new}^2 + k_{y,new}^2) + \mu - \Delta_{gap} \end{pmatrix} \quad (8)$$

where $k_{\pm} = k_{x,new} \pm ik_{y,new}$. The out of plane i.e. the z-directed magnetic field give rise to a Landau Gauge of $\vec{A} = (-By, 0, 0)$. In this case, the wave vector $k_y$ remains unchanged and $k_{x,new} = k_x + \frac{e}{\hbar}By$. The choice of Landau gauge may vary for any magnetic field. Here we have used the simplest form of Landau gauge for computational simplicity.

### C. Electrical Transport and Magnetization Dynamics

In general, the TI/FM bilayer heterostructures are small enough for applying the NEGF formalism of quantum transport [23]. Here, we use the standard self-consistent 2D NEGF method [23] to determine the total current and the spin current at the TI surface. Using our proposed Hamiltonian from the previous section, we first calculate the retarded Green's function defined as,

Retarded Green's Function, $G^R = [EI - H_{surface}^{eff}(k) - \Sigma_{c1} - \Sigma_{c2}]$ (9)

where $\Sigma_{c1}$ and $\Sigma_{c2}$ are the self-energy matrices for the contacts. Then using this retarded Green's function, the non-equilibrium Green's function can be written as:

Non equilibrium Green's Function, $G^n = G^R \Sigma^{in} G^A$ (10)

Here, $G^A$ is the complex conjugate of $G^R$ and $\Sigma^{in}$ is the strength of the contacts defined as: $\Sigma^{in} = [\Gamma_1]f_1 + [\Gamma_2]f_2$, where $f_1, f_2$ are the Fermi levels of the contacts and $\Gamma_1, \Gamma_2$ are defined as $\Gamma_1 = i[\Sigma_{c1} - \Sigma_{c1}^\dagger]$ and $\Gamma_2 = i[\Sigma_{c2} - \Sigma_{c2}^\dagger]$. Using these quantities we finally calculate the charge current and the spin current density as:

The total charge current density, $J = \frac{2\pi}{ih} \int Real[Trace\ (H_{surface}^{eff} G^n - G^n H_{surface}^{eff})]\ dE$ (11)

The spin current flowing in the TI surface [24], $J_{s0} = \frac{2\pi}{ih} \int Real[Trace\ (S.(H_{surface}^{eff} G^n - G^n H_{surface}^{eff}))]\ dE$ (12)

Where S is the Pauli spin matrix. This spin diffuses from the interface into the FM layer exerting torque on it [2].

Next, we apply the charge and spin currents to the LLG equation with Slonczewski spin-transfer torque term. The magnetization dynamics can be represented as [25] [26] [27]:

$$\frac{\partial \hat{m}}{\partial t} = -|\gamma|\hat{m} \times \vec{H_{eff}} + \alpha \hat{m} \times \frac{\partial \hat{m}}{\partial t} + \vec{Torque}$$ (13)

Here, $\vec{H_{eff}}$ is the effective magnetic field acting on the ferromagnet, $\hat{m}$ is the unit vector pointing to the direction of magnetization of the FM layer, $\alpha$ is the damping constant, $\gamma$ is the gyromagnetic ratio and $\vec{Torque}$ is the sum of field like torque and spin transfer like spin orbit torque [5] acting on the Ferromagnet. The field like torque can be calculated as $\vec{\tau}_{FL} = \Delta_{ex}\hat{m} \times \vec{n}_s$ [15], where $\vec{n}_s$ is the non-equilibrium spin density. This non-equilibrium spin density can be easily related to spin current density in the ferromagnetic layer as $J_s = -\mathcal{D}\vec{\nabla}.\vec{n}_s$ [28] where $\mathcal{D}$ is the diffusion coefficient of spin inside the FM layer.

Spin transfer like spin orbit torque is defined as the spatial change of the spin current. It can be expressed as [28]

$$\vec{\tau}_{ST} = \frac{1}{V} \iiint \left(-\vec{\nabla} J_s - \frac{1}{\tau_{sf}} \vec{n}_s\right) dV$$ (14)

Where V is the volume of the ferromagnet and $\tau_{sf}$ is the spin relaxation time. This spin relaxation compensates the spin current. According to fig. 2, spin diffuses in z direction into the FM layer. Therefore equation 14 can be simplified to

$$\vec{\tau}_{ST} = \frac{1}{d} \int_0^d \left( -\vec{\nabla}_z J_s - \frac{1}{\tau_{sf}} \vec{n}_s \right) dz \tag{15}$$

Where d is the thickness of the FM layer. Here torque consists of both in-plane and out of plane component. In order to solve this equation we need to get an expression for $\vec{n}_s$. If we consider the diffusion equation for the spin diffusion into the FM layer we can write during steady state [28]

$$\vec{\nabla} J_s = -\frac{1}{\tau_j} \vec{n}_s \times \hat{m} - \frac{1}{\tau_\phi} \hat{m} \times (\vec{n}_s \times \hat{m}) - \frac{1}{\tau_{sf}} \vec{n}_s \tag{16}$$

Where $\tau_j$ is the spin precession time and $\tau_\varphi$ is the spin decoherence time. The boundary conditions for solving this equation are $J_s(0) = pJ_{s0}$ where p is the spin injection efficiency from the interface into the FM layer and $J_s(d) = 0$. Also as we have stated earlier, spin diffuses in z direction. So we can assume that spin variation only exists along z-axis. Spin current at the interface has both in-plane and out-of-plane components i.e. $J_s(0) = J_{s,II}(0) + iJ_{s,\perp}(0)$. Using these conditions if we solve equation 16 we will get the following solution for the non-equilibrium spin density [28].

$$n_{s,II} + in_{s,\perp} = \left( C_1 exp\left(-\frac{z}{L}\right) + C_2 exp\left(\frac{z}{L}\right) \right) \tag{17}$$

Where L is defined as, $\frac{1}{L^2} = \frac{1}{\lambda_{sf}^2} + \frac{1}{\lambda_\phi^2} - \frac{i}{\lambda_j^2}$ [28]. $\lambda_{sf}, \lambda_j$ and $\lambda_\varphi$ are spin relaxation, spin precession and spin decoherence length inside the FM layer respectively. Using boundary conditions we get $C_1 = -\frac{n_{s0}exp(\frac{d}{L})}{exp(\frac{-d}{L})-exp(\frac{d}{L})}$

and $C_2 = \frac{n_{s0}exp(\frac{d}{L})}{exp(\frac{-d}{L})-exp(\frac{d}{L})}$. Here $n_{s0}$ is the non-equilibrium spin density at the bottom of FM layer (where it touches the TI). $n_{s0}$ can be easily related to spin current density at the interface by the equation, $n_{s0} = \frac{pLJ_{s0}}{D}$ [28]. Therefore eqn. (17) becomes

$$n_{s,II} + in_{s,\perp} = n_{s0}\left(\frac{\sinh\left(\frac{d-z}{L}\right)}{\sinh\left(\frac{d}{L}\right)}\right) = \frac{pJ_{s0}L}{D}\left(\frac{\sinh\left(\frac{d-z}{L}\right)}{\sinh\left(\frac{d}{L}\right)}\right) \tag{18}$$

Therefore the spin transfer torque becomes

$$\tau_{ST} = \frac{1}{d}\int_0^d \left(-\frac{n_{s0}D}{L^2}\frac{\sinh(\frac{d-z}{L})}{\sinh(\frac{d}{L})} - \frac{n_{s0}}{\tau_{sf}}\frac{\sinh(\frac{d-z}{L})}{\sinh(\frac{d}{L})}\right)dz \tag{19}$$

Diffusion coefficient is related to spin relaxation length by the Einstein-Smoluchowski equation as, $D\tau_{sf} = \lambda_{sf}^2$. Therefore the torque becomes

$$\tau_{ST,II} + i\tau_{ST,\perp} = \frac{pJ_{s0}L^2}{d}\left(\frac{1}{\lambda_\phi^2} - \frac{i}{\lambda_j^2}\right)\frac{\cosh\left(\frac{d}{L}\right)-1}{\sinh\left(\frac{d}{L}\right)} \tag{20}$$

In the TI/FM heterostructure, besides field-like spin orbit torque and spin-transfer like spin orbit torque, there is another important torque working in this heterostructure interface. This is called anti-damping torque and it appears due to Berry curvature [29]. It appears because the carrier spins experience two effective magnetic field due to FM layer magnetization and applied electric field. Because of these two effective B-fields, carrier spins become inclined towards z-axis and produce anti-damping torque. This torque can be incorporated in LLGS equation via the Gilbert damping constant, α. When the applied electric field and FM layer magnetization is perpendicular to each other there will be no anti-damping torque [29] and α is large as shown in our first simulated structure. But when the FM layer magnetization is not perpendicular to applied electric field, α becomes very small and the calculation for α in this case is shown later. Our calculation for α for both the cases agrees well with experimental observations [2][5].

### III. RESULTS AND DISCUSSION

In this section, we apply our proposed modeling framework on two different TI/FM heterostructures, the characteristics of which have recently been determined experimentally [2] [5] [8]. We first analyze the Cr doped $(Bi_{0.5}Sb_{0.5})_2Te_3$ / $(Bi_{0.5}Sb_{0.5})_2Te_3$ heterostructure and verify the validity of our results with experimental observations in ref. [5]. Then we check the consistency of our simulation framework with a Permalloy/$Bi_2Se_3$ heterostructure reported in ref. [2] [8].

## A. Cr doped (Bi$_{0.5}$Sb$_{0.5}$)$_2$Te$_3$ on top of (Bi$_{0.5}$Sb$_{0.5}$)$_2$Te$_3$

Recently it has been experimentally reported that if a TI is magnetically doped then it can behave like a magnet [30] [31] [32]. Therefore, it is possible to use a magnetically doped TI as the FM layer. The surface related issues is easily taken care of because the top TI layer has the same crystal structure and orientation as the bottom TI layer. However, the problem arises from the fact that the magnetically doped TI's Curie temperature is well below the room temperature. Experiments have been performed with Cr doped (Bi$_{0.5}$Sb$_{0.5}$)$_2$Te$_3$ / (Bi$_{0.5}$Sb$_{0.5}$)$_2$Te$_3$ heterostructure at very low temperature (1.9 K) [5]. Significant amount of spin transfer torque, strong enough to switch the top magnet, has been observed in this heterostructure. Here, the spin transfer torque is observed mainly along the z-axis. We first validate our model with this experiment.

### 1. 3D modeling of (Bi$_{0.5}$Sb$_{0.5}$)$_2$Te$_3$ and capturing the necessary effects with 2D surface modeling

In the experiment, a 6 nm thick Cr doped (Bi$_{0.5}$Sb$_{0.5}$)$_2$Te$_3$ quantum well is stacked over a 3 nm thick (Bi$_{0.5}$Sb$_{0.5}$)$_2$Te$_3$ quantum well [5]. In a very thin quantum well of TI, a bandgap opens up in the band diagram at the Dirac point due to a strong confinement effect [33]. Moreover, the presence of a strong magnetic exchange coupling further amplifies the bandgap opening [2]. Since the bandgap opening due to exchange coupling, $\Delta_{ex}$ cannot be determined experimentally, we first determine the bandgap opening, $\Delta_{conf}$ due to the quantum confinement and then gradually tune the $\Delta_{ex}$ parameter to match the conductivity.

In a 3nm thick quantum well of (Bi$_{0.5}$Sb$_{0.5}$)$_2$Te$_3$, the quantum confinement effect is extremely strong. Experimentally it has been observed that, due to surface state delocalization, Bi$_2$Te$_3$ and Sb$_2$Te$_3$ quantum wells have larger bandgap opening than Bi$_2$Se$_3$ [11] [12] [33]. Using the modified parameter set listed in Table I, we have calculated the bandgap opening of Bi$_2$Te$_3$ and Sb$_2$Te$_3$ quantum wells separately for different confinement lengths and compared with the experimental results [11] [12] (fig. 5). We can observe that, 3nm quantum wells of Bi$_2$Te$_3$ and Sb$_2$Te$_3$ have bandgap openings of 85meV and 35meV respectively. Using a mole fraction of 0.5, we can assume that the gap opening of 3nm thick (Bi$_{0.5}$Sb$_{0.5}$)$_2$Te$_3$ quantum well to be ~60meV (0.5*85 + 0.5*35).

We have also plotted the band diagram of a 3nm thick (Bi$_{0.5}$Sb$_{0.5}$)$_2$Te$_3$ quantum well (fig. 6) using Eqn. 1. The quantum confinement gap is $\sim 55\ meV$ near the Dirac cone which is very close to the above calculated value. This effect can be included in the 2D surface model in eqn. (8) through the parameter, $\Delta_{conf}$. In order to model this experiment, we use, $\Delta_{conf} = 30\ meV$ because this not only pulls up the conduction band by the amount of $\Delta_{conf}$ but also pulls the valence band down by the same amount.

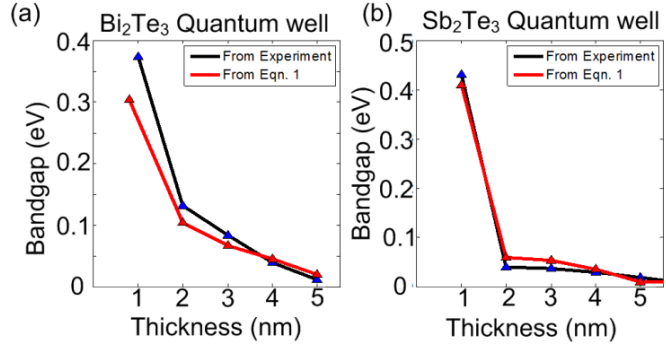

Fig. 5: Comparison of bandgap vs $(Bi_{0.5}Sb_{0.5})_2Te_3$ quantum well thickness between theoretical model using table I and experimental values [11] [12].

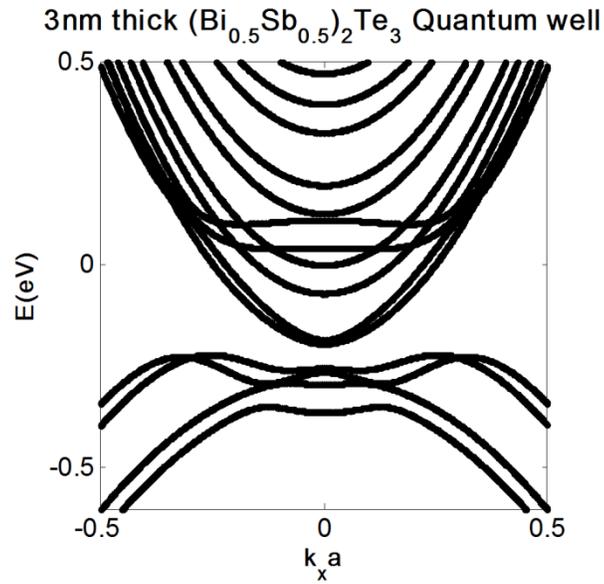

Fig. 6: E-k diagram of a 3nm thick $(Bi_{0.5}Sb_{0.5})_2Te_3$ quantum well. The well is confined in z-direction.

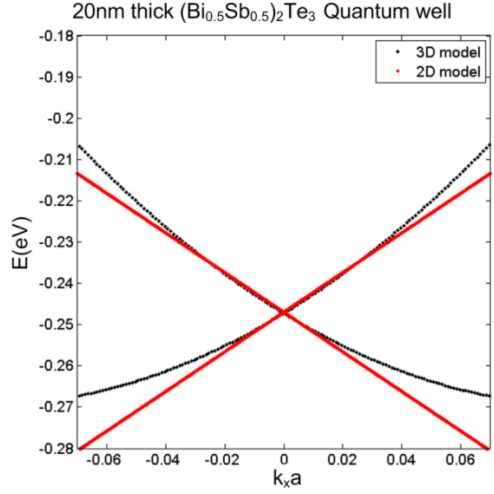

Fig. 7: Band diagram matching of 3D (Bi$_{0.5}$Sb$_{0.5}$)$_2$Te$_3$ quantum well and 2D surface modeling ('a' is the lattice constant).

Again in eqn. (8), the parameter A can be represented as, $A = \hbar v_F$ where $v_F$ is the Fermi velocity. Fermi velocity in Bi$_2$Te$_3$ and Sb$_2$Te$_3$ are measured as $\sim 3.5 \times 10^5\ ms^{-1}$ [11] and $\sim 4.3 \times 10^5\ ms^{-1}$ [34], respectively. Therefore, in (Bi$_{0.5}$Sb$_{0.5}$)$_2$Te$_3$, the Fermi velocity can be written as $0.5 \times (3.5 \times 10^5 + 4.3 \times 10^5) = 3.9 \times 10^5\ ms^{-1}$ (individual velocities multiplied by mole fractions and added) resulting in A = 2.6 eV-Å. To match the band diagrams, we have determined the fitting parameters, B1= B2 = 0.01 eV-Å². The Fermi level is calculated as, $\mu = -0.247\ eV$. Using these parameters we have calculated and matched the 3D band diagram of a 3nm thick z-confined (Bi$_{0.5}$Sb$_{0.5}$)$_2$Te$_3$ quantum well and the corresponding surface band diagram (free in x and y direction) as shown in fig. 7.

## 2. NEGF and LLG modeling and matching with the experiment

We have considered a large enough surface dimension in our model so that the quantum confinement along x and y axes does not create any noticeable effect on the band diagram. We found that a 100 nm × 100 nm surface dimension of (Bi$_{0.5}$Sb$_{0.5}$)$_2$Te$_3$ is large enough for this purpose. Using the 2D surface Hamiltonian with $\Delta_{ex} = 110\ meV$ and without considering the external magnetic field, we get a conductivity of $\sim 252\ S\ cm^{-1}$. It is very close to the experimentally calculated value of $\sim 219\ S\ cm^{-1}$. We observe a high out of plane spin Hall angle of 0.9425. One interesting observation is that, with the increment of TI thickness the spin Hall angle decreases gradually. Particularly for this structure, when the TI thickness is increased to 20 nm, the spin Hall angle decreases to 0.65.

In the experiment, an external magnetic field is applied and rotated in different directions for observing the effects [5] and finding the easy axis of the FM. It has been found that the easy axis is in the z-direction and an external magnetic field is applied in that direction. A z-directed external magnetic field not only shifts the band diagram of a TI but also creates a further bandgap opening equal to $(\frac{e}{\hbar}By)^2$ where y is the confinement length along y-axis and B is the external magnetic field. Therefore the conductivity further decreases in the presence of an external magnetic field due to the increment in the band gap. As we apply the LLGS equation to this structure using the currents calculated from the NEGF equation, we observe the magnetization dynamics with 1 V DC voltage as shown in fig. 8. The parameters for LLGS simulation are as follows:

TABLE II. Parameters for LLG simulation (simulated in MuMax3 [35]).

| Parameter | Value |
|---|---|
| Saturation Magnetization | $1.6 \times 10^6 \ A/m$ [5] |
| Exchange Constant | $1.9 \times 10^{-12} \ J/m$ [5] |
| Easy axis | Z axis [5] |
| uniaxial anisotropy constant | $7200 \ J/m^3$ [5] |
| External Magnetic Field | From -1.5 T to 1.5 T along Z axis |
| Out of plane Spin Hall angle | 0.94 |
| Gyromagnetic Ratio | $1.8 \times 10^{11} \ rad/sT$ [5] |
| Damping constant | 0.1 [5] |

We also observe a critical current of 8.5 μA for switching a 100 nm × 100 nm × 6 nm Cr doped $(Bi_{0.5}Sb_{0.5})_2Te_3$ magnet. The critical current density is $8.5\times10^4$ A cm$^{-2}$ which is very close to the experimentally reported value of $8.9\times10^4$ A cm$^{-2}$ [5]. Our simulations also suggest that the initial magnetization does not have noteworthy effect on the switching characteristics of the FM layer. In fig. 8(c), we can observe the switching of the magnet in the easy axis direction even in the absence of an external magnetic field.

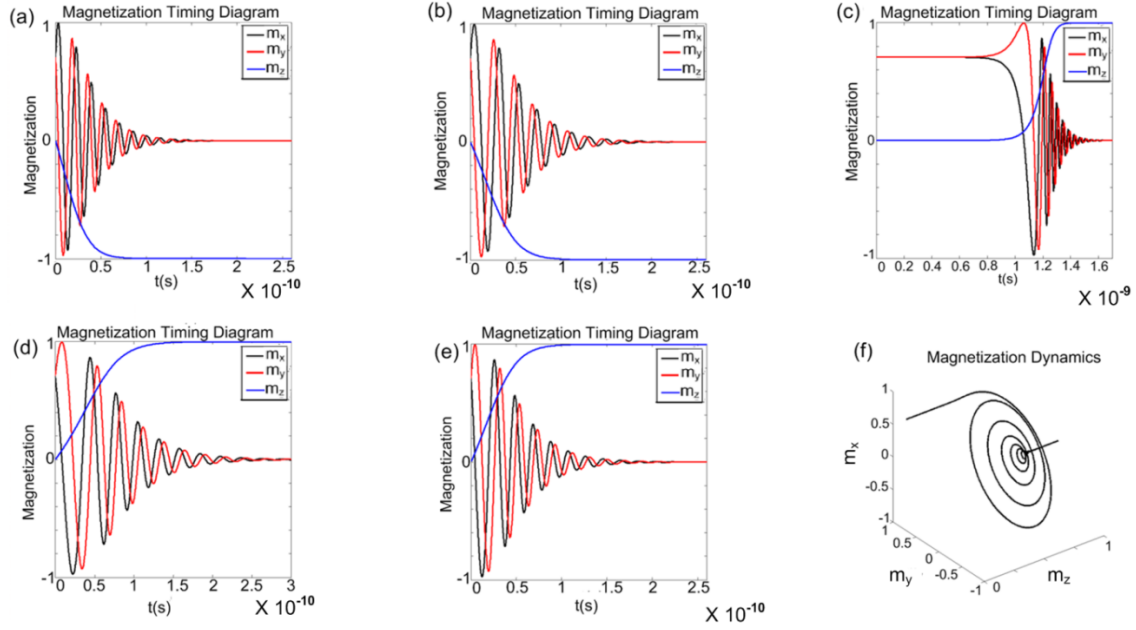

Fig. 8: Magnetization components as a function of time for 1 V DC voltage and z-directed external magnetic field of (a) -1.5 T, (b) -1 T, (c) 0 T, (d) 0.5 T, (e) 1 T. (f) Changes in different magnetization components $m_x$, $m_y$ and $m_z$ for no external magnetic field.

### B. Permalloy on top of $Bi_2Se_3$

As we have discussed above, the Cr doped $(Bi_{0.5}Sb_{0.5})_2Te_3$ has a Curie temperature that is below the room temperature. For room temperature operations, a Permalloy can be coupled with the $Bi_2Se_3$ and experiments have been conducted [2][8] to calculate the spin current conductivity defined as $\sigma_{s,i} = \frac{J_{s,i}}{E}$ where $J_{s,i}$ is the i-th component of the spin current density and E is the electric field. Here, the spin accumulation is observed mainly in the x-y plane exerting a torque on the Permalloy and thus changing the magnetization direction of the FM layer.

#### 1. 3D modeling of $Bi_2Se_3$ and capturing the necessary effects with 2D surface modeling

In the experiment, 8 nm and 16 nm thick quantum wells of $Bi_2Se_3$ were used [2]. If we consider the quantum confinement effect in this case, the bandgap at Dirac point is negligible (~0.3 meV for 8 nm thick well and ~0.7 μeV for 16 nm thick well). In fig. 9, we have shown the effect of quantum confinement in $Bi_2Se_3$ for different quantum well thickness and it is evident that the confinement effect is prominent for thickness less than or equal to 6 nm [33].

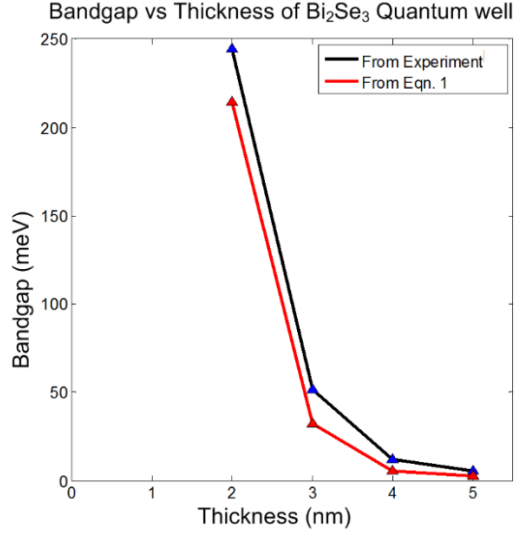

Fig. 9: Comparison of $Bi_2Se_3$ quantum well thickness between theoretical model and experimental values [33].

Therefore, for 8 nm and 16 nm $Bi_2Se_3$ quantum wells, we take, $\Delta_{conf}= 0$. From eqn. (1), the E-k diagrams of 8 nm and 16 nm quantum wells of $Bi_2Se_3$ are calculated and shown in fig. 10 which also shows no significant bandgap opening. For matching the band structure with 2D surface modeling, we have calculated the fitting parameter A to be 1.69 eV-Å in eqn. (8). Here the conduction and valence bands are not symmetric. Therefore, the other fitting parameters, B1 and B2 have separate values and for matching with the 3D band diagram, their values are calculated to be 5 eV-Å² and 9 eV-Å². The Fermi level is calculated to be, $\mu = -0.225\ eV$. Using these parameters, the comparison of 3D quantum well and 2D surface band diagram is shown in fig. 11. Here we have matched the band diagram around the energy of $V_d q \pm 3kT$ of the Dirac point, where $V_d$ is the applied voltage, q is the electron charge, T is the temperature and k is the Boltzmann constant. This is appropriate since we are focusing on the spin transport properties of the TI/FM heterostructure and current flow mainly takes place within the range of $V_d q \pm 3kT$ around the Fermi level. Here, 0.2 V DC applied to the contacts in fig. 2 is large enough to operate this heterostructure and to switch the magnet.

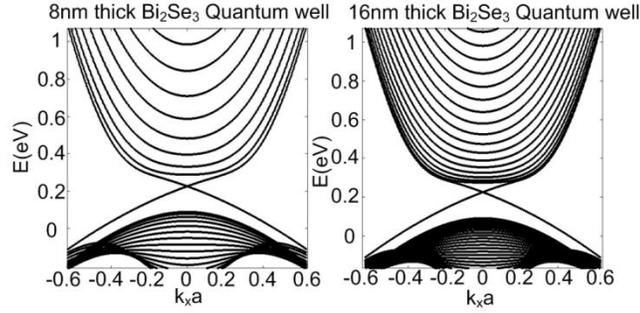

Fig. 10: E-k diagrams of 8nm and 16nm thick $Bi_2Se_3$ quantum wells. The well is confined in the z-direction.

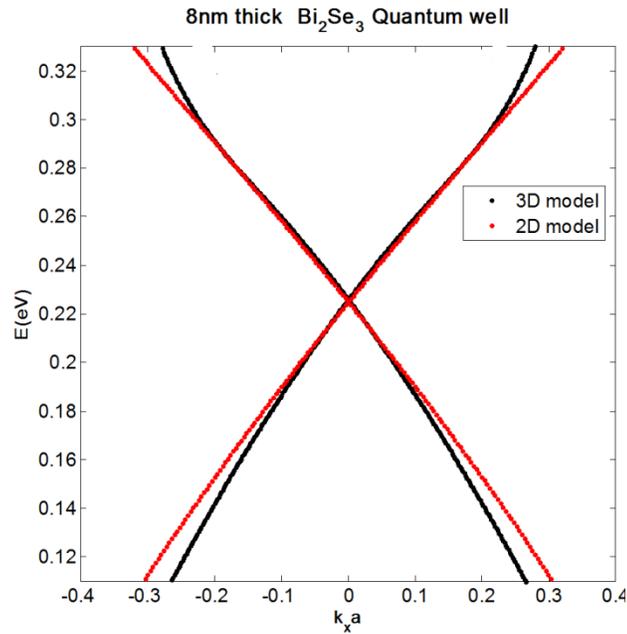

Fig. 11: Band diagram matching of 3D $Bi_2Se_3$ and 2D surface modeling.

## 2. NEGF and LLG modeling and matching with the experimental data

In the experiments [2] [8], the external magnetic field is applied in-plane i.e., in x-y plane and therefore it shifts the band diagram by the amount of $\frac{e}{\hbar}Bz$ and creates a bandgap equal to $(\frac{e}{\hbar}Bz)^2$ where z is the confinement length along z-axis and B is the external magnetic field. However, the external magnetic field used in the experiment is very small (only to cancel the stray fields) and therefore its effect can easily be ignored in this case. Again, there is a strong exchange coupling between $Bi_2Se_3$ and Permalloy. But as the magnetization is in-plane, there will be no bandgap opening, hence $\Delta_{ex}=$ 0 meV. Using our model, we calculate the charge current conductivity and in-plane and out-of-plane spin conductivity as shown in Table III:

TABLE III. Conductivity comparison between theoretical model and experimental measurement

| Conductivity | Calculation from theoretical model (Eq. 5) $\Omega^{-1}m^{-1}$ | Practical measurement [8] $\Omega^{-1}m^{-1}$ |
|---|---|---|
| Charge current conductivity | $9.5 \times 10^4$ | $5.7 \times 10^4$ |
| In plane spin current conductivity (8nm $Bi_2Se_3$) | $5.4 \times 10^4 \frac{\hbar}{2e}$ | $3 \times 10^4 \frac{\hbar}{2e}$ |
| Out of plane spin current conductivity (8nm $Bi_2Se_3$) | $4.16 \times 10^4 \frac{\hbar}{2e}$ | $2.5 \times 10^4 \frac{\hbar}{2e}$ |
| In plane spin current conductivity (16nm $Bi_2Se_3$) | $9.67 \times 10^4 \frac{\hbar}{2e}$ | $5.5 \times 10^4 \frac{\hbar}{2e}$ |
| Out of plane spin current conductivity (16nm $Bi_2Se_3$) | $1.16 \times 10^5 \frac{\hbar}{2e}$ | $6.7 \times 10^4 \frac{\hbar}{2e}$ |

In order to make the band structure free from quantum confinement in the x and the y directions, we have considered a 60 nm × 60 nm surface dimension which is large enough for this purpose (observed by calculating the Eigen values for this structure). A very important figure of merit is the spin Hall angle which is defined as $\frac{2q}{\hbar}\frac{J_s}{J_T}$, where $J_s$ is the spin current density and $J_T$ is the total charge current density [8]. We have observed a high in plane spin Hall angle of 1.1 for 16 nm thick $Bi_2Se_3$, matching well with experimentally observed value (~1.00 [8]). We also have found that for 16 nm thick $Bi_2Se_3$, the out of plane spin hall angle is 1.03 while it is experimentally reported as ~1.1 [8].

Another important factor is the calculation of spin transfer torque acting on the FM layer. Let us consider the spin injection efficiency at the interface, p = 1. Now in Permalloy spin diffusion length, $\lambda_{sf}$ = 5nm [2]. Assuming realistic value for spin decoherence length ($\lambda_\varphi = 1nm$ [2]) and spin precession length ($\lambda_j = 1nm$ [2]) we have calculated the in plane torque to be $1.84 \times 10^{-5}$ T from equation (20) while it has been experimentally determined as $2.7 \times 10^{-5}$ T.

For applying the LLG equation, we first determine the damping factor due to the absence of experimentally observed value in this respect. From the theory of ST-FMR experiment [2], the damping factor is defined as $\alpha = \frac{\mu\gamma\Delta H}{4\pi f'}$ [36] where f' is the frequency which is used to measure the swept field linewidth, $\Delta H$ is the linewidth and $\gamma$ is the gyromagnetic ratio. In this experiment, when frequency f' = 8 GHz the linewidth, $\Delta H = 1.94\ mT$. Substituting, $\mu\gamma = 2.8\ GHz/KOe$ we get the damping constant, $\alpha = 5.4 \times 10^{-4}$, which is comparable with typically observed value of damping constant for similar type of structures [37]. In order to verify the validity of this parameter we have applied an ac current of 7.7 mA at a frequency of 8 GHz as done in the experiment [2] and used the following parameters in LLG equation:

TABLE IV. Parameters for LLG simulation.

| Parameter | Value |
| --- | --- |
| Saturation Magnetization | $0.86 \times 10^6\ A/m$ [38] |
| Exchange Constant | $1.3 \times 10^{-11}\ J/m$ [38] |
| Easy axis | $45^0$ with X axis in X-Y plane [2] |
| uniaxial anisotropy constant | $500\ J/m^3$ [39] |
| External Magnetic Field | From 0.05 T to 0.09 T along easy axis |
| In plane Spin Hall angle | 1.1 |
| Out of plane Spin Hall angle | 1.03 |
| Damping constant | $5.4 \times 10^{-4}$ |

Then we have observed oscillations in magnetization which is consistent with the experimental observations (fig. 12). The initial magnetization direction of the Permalloy is assumed to be in the z-direction.

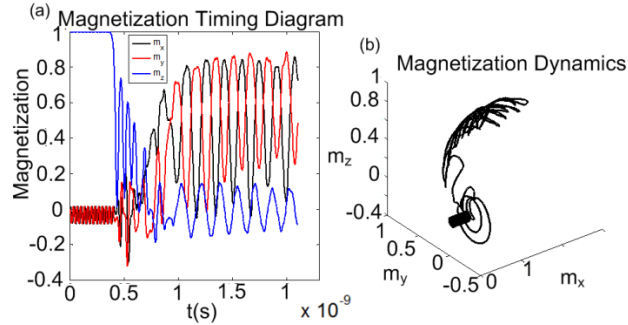

Fig. 12: Magnetization dynamics observed after applying ac current of 7 mA at a frequency of 8 GHz.

The main hindrance in this structure appears to be the amount of consumed power. As Permalloy is a ferromagnetic metal and therefore, its conductivity is high. On the other hand, in this heterostructure the conductivity of $Bi_2Se_3$ lowers due to the bandgap opening at its Dirac point. As a result, for the structure shown in fig. 2, large amount of current shunts through the adjacent Permalloy. Experimental results suggest that the current shunting through the Permalloy is 25 times higher than the current flowing through $Bi_2Se_3$ surface. Applying the NEGF equation at a dc voltage of 0.25 V, the total power consumption is calculated to be $\sim 2.31\ mW$. Therefore, if this structure is used as a single cell in a memory array, a memory of only 500 bits will require almost 1.15 W of power which is orders of magnitude higher than the current memory structures.

## C. Magnetic memory using TI/Ferromagnet heterostructure

Traditional Magnetic Tunnel Junction (MTJ) based memory designs require comparatively high operating voltages due to the necessity of large tunneling current [40]. Moreover, large access transistor are used for driving this current and reliability issues relating to high tunneling current conduction through MgO have appeared as serious drawbacks for applying traditional MTJs in conjunction with traditional CMOS technology. Hence, it is prudent to explore TI/FM bilayer heterostructures as solution to these problems. For such structures, the critical current required to switch the magnet is small especially for Cr doped $(Bi_{0.5}Sb_{0.5})_2Te_3$ / $(Bi_{0.5}Sb_{0.5})_2Te_3$ heterostructure. The small critical current would allow us to operate the device at a very low voltage (~0.2 V). In fig. 13(a), we propose a memory cell using TI/Ferromagnet bilayer heterostructure. This memory cell has three layers. The first layer is the MTJ with TI at the bottom for switching the free magnetic layer. The write current flows through the TI, and depending on the direction of the current, the free layer is expected to switch, leading to a change in the tunneling magnetoresistance, which can interpreted as binary '1' and '0', accordingly. The second layer is the antiferromagnetic (AFM) layer made of Ni-Mn. It can also be made of Pt-Mn and Tb-Mn depending on the fabrication feasibility and application. This layer is used to stabilize the magnetization of fixed magnet. The third layer comprises of Ruthenium (Ru) and CoFe between the AFM and the MTJ. This layer is called the synthetic antiferromagnetic (SAF) layer and used to fix the magnetization of the fixed magnetic layer by cancelling the stray fields around it [41]. In the proposed device, the CoFe, through nonmagnetic material Ru, is exchange coupled to the free magnetic layer [42] and Ru provides a strong Ruderman-Kittel-Kasuya-Yosida (RKKY) interlayer coupling [41]. In fig. 13(b), we show a single memory cell along with the read write circuit. The top terminal is connected to read bit line $BL_{read}$. When read current passes from the top to bottom as shown in figure, we can sense the tunneling magnetoresistance depending on the magnetization direction of free and fixed magnetic layer. One of the other terminals is connected to one write bit line and another terminal is connected to another write line via a pass transistor. Two write lines are used to allow current flow in both directions to get alternate magnetization switching. The pass transistor is controlled via the word line (WL).

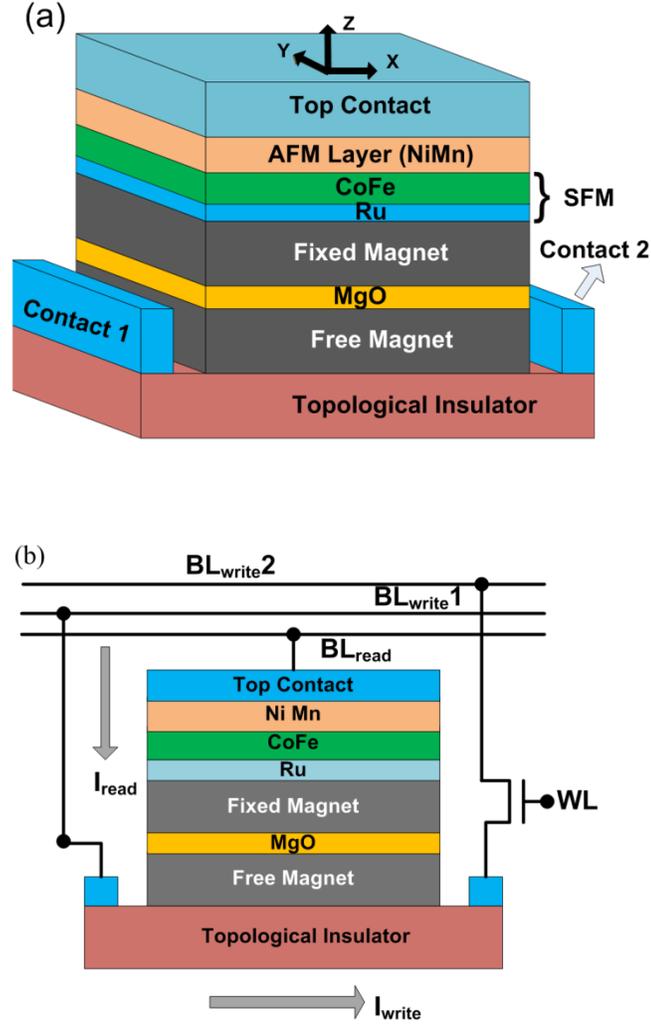

Fig. 13: (a) 3D diagram of the proposed memory design. (b) One memory bit cell with related circuit.

We observed large write power consumption in the Permalloy/$Bi_2Se_3$ heterostructure mainly due to the shunting current as stated in section III-B. As the conductivity of Permalloy is orders of magnitude higher than that of $Bi_2Se_3$, it is difficult to reduce the ratio of shunting current loss. Therefore, from a practical point of view, this structure may not be promising for memory design. Let us consider the Cr doped $(Bi_{0.5}Sb_{0.5})_2Te_3$ / $(Bi_{0.5}Sb_{0.5})_2Te_3$ structure. In the experiment, a 6nm thick Cr doped $(Bi_{0.5}Sb_{0.5})_2Te_3$ quantum well (QW) was used as the Ferromagnet and a 3nm thick $(Bi_{0.5}Sb_{0.5})_2Te_3$ QW well was used as the TI [5]. For such a device, almost half of the current shunts through the top magnetic layer. Now, for our device simulations, to reduce the conductivity of the free layer, the thickness of Cr doped $(Bi_{0.5}Sb_{0.5})_2Te_3$ FM layer is reduced to 3nm which was 6 nm in the experiment. Similarly, for increasing the conductivity of the bottom TI layer we have increased the thickness of $(Bi_{0.5}Sb_{0.5})_2Te_3$ TI layer to 20 nm which was 3 nm in the experiment, so that, $\Delta_{conf} = 0\ eV$. While confining in x and y directions, we have to make sure that it does not introduce any confinement gap opening at the Dirac point and we have found that a 100 nm ×100 nm x-y surface is large enough for this purpose. For our memory bit-cell simulations, the magnet size is assumed to be 100 nm × 100 nm × 3 nm with the TI size to be 100 nm × 100 nm × 20 nm. With such a structure, the amount of shunting current is only 5%-6% of the total current. As we have mentioned above, 0.2 V DC applied to the contacts as shown in fig. 2 is sufficient for switching this structure. For read operation, since the read current flows through the MTJ, and is comparatively smaller, the power consumption during the read operation is small. Major power consumption takes place during the write operation. If we operate this structure at 0.2 V, one cell consumes 20 μW of power and it takes ~6 ns to switch. The timing diagram and magnetization diagram is shown in fig. 14a and 14b, respectively.

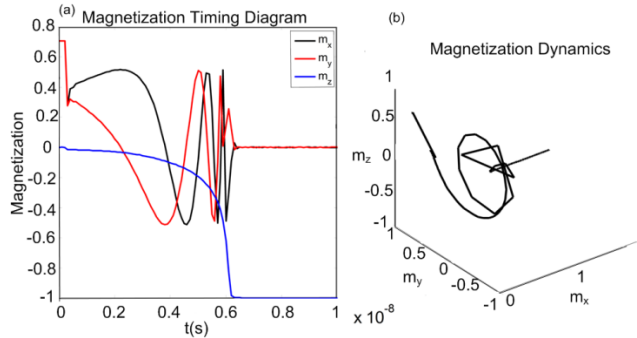

Fig. 14: (a) Magnetization Timing diagram and (b) magnetization dynamics of Cr doped $(Bi_{0.5}Sb_{0.5})_2Te_3$ / $(Bi_{0.5}Sb_{0.5})_2Te_3$ heterostructure based memory structure of fig. 13.

The main drawback of this memory structure is that it operates at low temperatures. This issue can be solved by using a ferromagnetic insulator as free magnet that has a high Curie temperature. It will also minimize the amount of shunting current and hence, the power consumption will be reduced.

## IV. CONCLUSION

To summarize, we have developed a modeling and simulation framework for TI/FM heterostructure which includes proposing a TI surface Hamiltonian that captures all necessary features of TI surface states for spin transport calculations. We have associated this Hamiltonian with self-consistent NEGF formalism to determine the charge and the spin transport. Finally, the magnetization dynamics (using LLGS) is used to observe the magnetization timing diagram of the FM layer. We have validated our model by benchmarking against experimental results on TI/FM heterostructures. Our proposed simulation framework is computationally efficient because of using a small 2×2 surface Hamiltonian and yet it produces good results. We have applied our model to analyze a memory cell using TI/FM bilayer heterostructures. Our results indicate that if the shunting current through the ferromagnet can be reduced, the memory cell can be energy-efficient as it can be operated at low voltages due to high charge current to spin current conversion ratio. We have also shown a way of reducing this shunting current in Cr doped $(Bi_{0.5}Sb_{0.5})_2Te_3$ / $(Bi_{0.5}Sb_{0.5})_2Te_3$ heterostructure. However, such a device can only operate at very low temperature. If the shunting current can be reduced and Curie temperature of magnetically doped TI can be increased, TI based memory device can be promising because of its simplicity, compact area and non-volatility.

## ACKNOLEDGEMENT

This research was funded in part by C-SPIN, the center for spintronic materials, interfaces, and architecture, funded by DARPA and MARCO; the Semiconductor Research Corporation, the National Science Foundation, and the NSSEFF program.

## REFERENCES

[1] H. Zhang, C. X. Liu, X. L. Qi, X. Dai, Z. Fang, and S. C. Zhang, "Topological insulators in $Bi_2Se_3$, $Bi_2Te_3$ and $Sb_2Te_3$ with a single Dirac cone on the surface," Nat. Phys. 5, 438 (2009).

[2] A. R. Mellnik, J. S. Lee, A. Richardella, J. L. Grab, P. J. Mintun, M. H. Fischer, A. Vaezi, A. Manchon, E. A. Kim, N. Samarth et al., "Spin-transfer torque generated by a topological insulator," Nature 511, 449 (2014).

[3] Edelstein, V. M, "Spin polarization of conduction electrons induced by electric current in two-dimensional asymmetric electron systems," Solid State Commun.73, 233–235 (1990).

[4] Seokmin Hong, Vinh Diep, and Supriyo Datta, "Modeling potentiometric measurements in topological insulators including parallel channels," Phys. Rev. B. 86, 085131 (2012).

[5] Fan, Y. et al., "Magnetization switching through giant spin-orbit torque in a magnetically doped topological insulator heterostructure," Nature Mater, 13, 699–704 (2014).

[6] Y. Lu and J. Guo, "Quantum simulation of topological insulator based spin transfer torque device," Appl. Phys. Lett., vol. 102, p. 073106, (2013).

[7] Liu C-X, Qi X-L, Zhang H, Dai X, Fang Z and Zhang S.C, "Model Hamiltonian for topological insulators," Phys. Rev. B 82 045122, (2010).

[8] Alex Ryckman Mellnik, Ph.D. dissertation, Cornell University, (2015).

[9] P. Yu, M. Cardona, *Fundamentals of Semiconductors: Physics and Materials Properties*, 3rd ed., (Springer. Section 2.6, 2005) p. 68.

[10] M. Ehrhardt and T. Koprucki (eds.), *Multi-Band Effective Mass Approximations: Advanced Mathematical Models and Numerical Techniques,* Lecture Notes in Computational Science and Engineering 94, (Springer. Section 6.2.4, 2014) p. 227.

[11] Li, Y.-Y. et al. "Growth dynamics and thickness-dependent electronic structure of topological insulator Bi2Te3 thin films on Si,". Preprint at http://arxiv.org/abs/ 0912.5054v1 (2009).

[12] G. Wang, X. Zhu, J. Wen, X. Chen, K. He, L. Wang, X. Ma, Y. Liu, X. Dai, Z. Fang, J. Jia, and Q. Xue, "Atomically Smooth Ultrathin Films of Topological Insulator $Sb_2Te_3$," Nano Res. 3, 874 (2010).

[13] C. L. Kane and E. J. Mele, "Z2 Topological Order and the Quantum Spin Hall Effect," Phys. Rev. Lett. 95, 146802 (2005).

[14] Rui Yu et al. "Quantized Anomalous Hall Effect in Magnetic Topological Insulators**,"** Science 329, 61 (2010).

[15] T. Yokoyama, J. Zang, and N. Nagaosa, "Theoretical study of the dynamics of magnetization on the topological surface," Phys. Rev. B 81, 241410 (2010).

[16] Qi, X.-L., Hughes, T. L. & Zhang, S.-C. "Topological field theory of time-reversal invariant insulators", Phys. Rev. B 78, 195424 (2008)

[17] H.-Z. Lu, J. Shi, and S.-Q. Shen," Competition between Weak Localization and Antilocalization in Topological Surface States", Phys. Rev. Lett. 107, 076801 (2011).

[18] Z. Jiang, F. Katmis, C. Tang, P. Wei, J. S. Moodera, and J. Shi," A comparative transport study of Bi2Se3 and Bi2Se3/yttrium iron garnet", Appl. Phys. Lett. 104, 222409 (2014).

[19] Sinova, J. *et al.*," Universal Intrinsic Spin Hall Effect", Phys. Rev. Lett. **92,** 126603 (2004).

[20] S. Narishige, K. Mitsuoka, and Y. Sugita, "Crystal Structure and Magnetic Properties of Permalloy Films Sputtered by Mixed," IEEE Trans. Magn. 28, 990 (1992).

[21] Zhang, W., Yu, R., Zhang, H.-J., Xi, D. & Fang, Z. "First principles studies on 3-dimensional strong topological insulators: Bi2Se3, Bi2Te3 and Sb2Te3." Preprint at http://arxiv.org/ abs/cond-mat/1003.5082 (2010).

[22] J.M. Luttinger, "The Effect of a Magnetic Field on Electrons in a Periodic Potential," Phys. Rev. B 84 (1951) 814.

[23] S. Datta, *Quantum Transport : Atom to Transistor*, 2nd ed., (Cambridge University Press (2005)).

[24] Charles Augustine, Arijit Raychowdhury, Dinesh Somasekhar, James Tschanz, Vivek De and Kaushik Roy, "Numerical Analysis of Typical STT-MTJ Stacks for 1T-1R Memory Arrays**,"** IEEE Transactions on Electron Devices, Vol. 58, No. 12, December, (2011).

[25] Xuanyao Fong, Ph.D. dissertation, Purdue University, (2014).

[26] Massimiliano d'Aquino, PhD thesis, Universit`a degli studi di Napoli "Frederico II", (2004).

[27] J.C. Slonczewski, "Current-driven excitation of magnetic multilayers", Journal of Magnetism and Magnetic Materials, 159 (1996), L1-L7.

[28] A. Manchon, R. Matsumoto, H. Jaffres, and J. Grollier," Spin transfer torque with spin diffusion in magnetic tunnel junctions", Phys. Rev. B 86, 060404 (2012).

[29] H. Kurebayashi, Jairo Sinova, D. Fang, A. C. Irvine, T. D. Skinner, J. Wunderlich, V. Novak, R. P. Campion, B. L. Gallagher, E. K. Vehstedt, L. P. Zarbo, K. Vyborny, A. J. Ferguson and T. Jungwirth, "An antidamping spin–orbit torque originating from the Berry curvature", Nat. Nanotechnol. 9, 211 (2014).

[30] Kou, X. et al., "Manipulating surface-related ferromagnetism in modulation-doped topological insulators," Nano Lett. 13, 4587–4593 (2013).

[31] Kou, X. F. et al., "Magnetically doped semiconducting topological insulators," J. Appl. Phys. 112, 063912 (2012).

[32] Checkelsky, J. G., Ye, J., Onose, Y., Iwasa, Y. & Tokura, Y. Dirac-fermion-mediated ferromagnetism in a topological insulator. Nature Phys. 8, 729–733 (2012).

[33] Chang , C . Z . et al., "Growth of quantum well films of topological insulator $Bi_2Se_3$ on insulating substrate," SPIN 1, 21 – 25 (2011).

[34] Y. Jiang, Y. Wang, M. Chen, Z. Li, C. Song, K. He, L. Wang, X. Chen, X. Ma, and Q. K. Xue, "Landau Quantization and the Thickness Limit of Topological Insulator Thin Films of Sb2Te3," Phys. Rev. Lett. 108, 016401 (2012).

[35] A. Vansteenkiste et al., "The design and verification of MuMax3," AIP Adv. 4, 107133 (2014).

[36] Patton, Carl E. and Nan Mo, *Appendix F: FMR Linewidth Measurements*, (Colorado State University).

[37] B. Heinrich, C. Burrowes, E. Montoya, B. Kardasz, E. Girt, Y.-Y. Song, Y. Sun, and M. Wu, "Spin Pumping at the Magnetic Insulator (YIG)/Normal Metal (Au) Interfaces," Phys. Rev. Lett. 107, 066604 (2011)

[38] Jinghua Guo (Ed.), "*X-Rays in Nanoscience: Spectroscopy, Spectromicroscopy, and Scattering Techniques,*" (Wiley, New York, 2010) p. 71.

[39] Hertel, R., "Thickness dependence of the magnetization structures of thin permalloy rectangles", Z. Metallkd. 93, 957–962. (2002).

[40] Sasikanth Manipatruni, Dmitri E. Nikonov, Ian A. Young, "Voltage and Energy-Delay Performance of Giant Spin Hall Effect Switching for Magnetic Memory and Logic", arXiv:1301.5374v1 [cond-mat.mes-hall]) (Submitted on 23 Jan, 2013).

[41] K.H.J. Buschow (Ed.), *Handbook of Magnetic Material*, Vol. 19, (Elsevier North-Holland, Amsterdam, 2011) p. 71.

[42] D. D. Tang and Y.J. Lee, *Magnetic Memory: Fundamentals and Technology*, 1st edn., (Cambridge University Press, New York, 2010) p. 109.